\documentclass[conference]{IEEEtran}
\IEEEoverridecommandlockouts

\usepackage{algorithm}

\usepackage{multirow}

\usepackage{booktabs} 

\usepackage{cite}
\usepackage{amsmath,amssymb,amsfonts}
\usepackage{algorithmic}
\usepackage{graphicx}
\usepackage{textcomp}
\usepackage{xcolor}
\def\BibTeX{{\rm B\kern-.05em{\sc i\kern-.025em b}\kern-.08em
    T\kern-.1667em\lower.7ex\hbox{E}\kern-.125emX}}
\begin{document}

\title{Adversarial Infrared Curves: An Attack on Infrared Pedestrian Detectors in the Physical World

}


\author{\IEEEauthorblockN{1\textsuperscript{st} Chengyin Hu  \quad 2\textsuperscript{st} Weiwen Shi}
\IEEEauthorblockA{\textit{School of Computer Science and Engineering} \\
\textit{University of Electronic Science and Technology of China}\\
chengdu, China \\
cyhuuestc@gmail.com \quad weiwen\_shi@foxmail.com}}

\maketitle


\begin{abstract}

Deep neural network security is a persistent concern, with considerable research on visible light physical attacks but limited exploration in the infrared domain. Existing approaches, like white-box infrared attacks using bulb boards and QR suits, lack realism and stealthiness. Meanwhile, black-box methods with cold and hot patches often struggle to ensure robustness.
To bridge these gaps, we propose Adversarial Infrared Curves (AdvIC). Using Particle Swarm Optimization, we optimize two Bezier curves and employ cold patches in the physical realm to introduce perturbations, creating infrared curve patterns for physical sample generation. Our extensive experiments confirm AdvIC's effectiveness, achieving 94.83\% and 67.2\% attack success rates for digital and physical attacks, respectively. Stealthiness is demonstrated through a comparative analysis, and robustness assessments reveal AdvIC's superiority over baseline methods.
When deployed against diverse advanced detectors, AdvIC achieves an average attack success rate of 76.84\%, emphasizing its robust nature. We conduct thorough experimental analyses, including ablation experiments, transfer attacks, and adversarial defense investigations. Given AdvIC's substantial security implications for real-world vision-based applications, urgent attention and mitigation efforts are warranted.
\end{abstract}


\begin{figure*}[ht]
\centering
\includegraphics[width=1\linewidth]{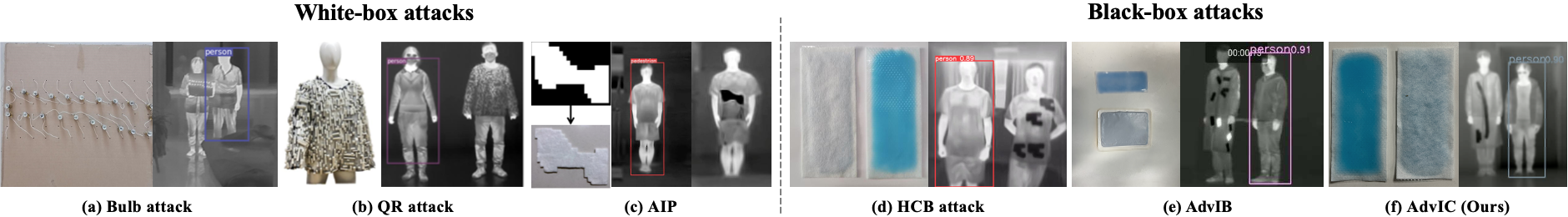} 
\caption{Demonstration of our proposed AdvIC and other infrared physical attacks.}
\vspace{-0.3cm}
\label{figure1}
\end{figure*}

\section{Introduction}

Currently, deep neural networks (DNNs) have demonstrated notable achievements across diverse domains, including image recognition \cite{ref1}, speech recognition \cite{ref2}, natural language processing \cite{ref3}, and object detection \cite{ref4}. As a pivotal technology in artificial intelligence, DNNs play a crucial role in advancing various applications.

In the context of object detection, some studies \cite{ref42,ref43} emphasize training robust detectors, specifically in the visible light spectrum, by curating datasets in visible light environments. Recognizing the limitations of visible light detectors in low-light conditions, some researchers \cite{ref40} have progressively shifted their focus to the infrared domain. This entails collecting infrared samples and training detectors specialized for the infrared spectrum.

While DNNs-based tasks have achieved significant progress, their security has become a focal point for many scholars \cite{ref5,ref6}. Presently, most studies \cite{ref7,ref8} are dedicated to examining attacks in digital environments by introducing subtle perturbations to deceive advanced DNNs, typically imperceptible to human observers. Other works \cite{ref9,ref10} concentrate on attacks in the physical realm, employing physical perturbations on the target object's surface. Samples are then collected through sensors and input into DNN-based systems to disrupt their functionality. Physical attacks differ from digital attacks, as adversarial perturbations must be designed to be perceptible by sensors. In the realm of physical attacks on object detectors, current efforts \cite{ref11,ref12} predominantly focus on visible light detectors, utilizing poster pasting to alter the texture of the target object. In contrast, some recent studies \cite{ref13,ref14} propose physical attacks against infrared detectors. In these attacks, infrared perturbations are designed to modify the temperature of the target object's region using heating and cooling materials. The resulting patterns of white and black on the infrared camera correspond to the hot and cool materials, respectively. Leveraging this property, attackers strategically design and optimize the position and shape of these materials to execute effective attacks on infrared detectors.

As depicted in Figure \ref{figure1}, certain studies employ bulb board \cite{ref13} (refer to Figure \ref{figure1} (a)), QR suits \cite{ref14} (refer to Figure \ref{figure1} (b)), and aerogel patches \cite{ref15} (refer to Figure \ref{figure1} (c)) as physical perturbations for executing white-box attacks against infrared pedestrian detectors. However, these approaches exhibit two notable drawbacks. Firstly, they lack stealthiness, deploying perturbations outside the clothing of pedestrians, making them prone to arousing suspicion. Secondly, white-box attacks often struggle to align with realistic scenarios. This is attributed to the difficulty for attackers to obtain internal information about the target model. Other works utilize cold and hot patches \cite{ref16,ref17} (refer to Figure \ref{figure1} (d) and (e)) as physical perturbations for conducting black-box attacks on object detectors. Nevertheless, the robustness of these methodologies remains uncertain.

Currently, the challenge in conducting physical attacks on infrared detectors lies in finding a balance between stealthiness and robustness. Small perturbations, while capable of achieving stealthiness, often fall short in terms of robustness. On the other hand, large perturbations, which enhance robustness, usually come at the cost of reduced stealthiness.

Building on the preceding discussion, we introduce a novel physical attack against infrared pedestrian detectors, named Adversarial Infrared Curves (AdvIC). This approach involves emulating Bezier curves and employing PSO \cite{ref38} for optimization, ultimately yielding the most adversarial Bezier curves. In the physical realm, cold patches are strategically placed inside the pedestrian's clothing to generate stealthy physical samples. In comparison to previous methods such as the bulb attack \cite{ref13}, QR attack \cite{ref14}, and AIP \cite{ref15}, our approach excels in stealthiness as the perturbation is concealed within the clothing, making it challenging to detect without an infrared camera. Additionally, compared to HCB \cite{ref16} and AdvIB \cite{ref17}, our method proves to be more efficient in perturbation deployment and attack execution, demonstrating superior robustness as verified in the experimental results detailed in Section \ref{sec4}. Notably, the deployment cost of our method is economical, with a total cost not exceeding \$5, rendering it highly accessible for implementation. Our primary contributions can be summarized as follows:

(1) We employ Bezier curves in the development of a novel black-box physical attack, AdvIC, targeting infrared pedestrian detectors. This approach achieves a balanced tradeoff between stealthiness and robustness. Notably, our method is characterized by easy deployment and cost-effectiveness, presenting a significant security threat in physical scenarios.

(2) A comprehensive series of experiments is conducted to validate the effectiveness, stealthiness, and robustness of the proposed method across both digital and physical environments. The experimental results demonstrate a higher attack success rate for our method compared to the baseline.

(3) In-depth analysis of our proposed method includes shape ablation experiments, confirming the superiority of our curve design over alternative shapes (line, triangle, circle, etc.). Transfer attack experiments are conducted to assess the method's transferability. Additionally, we anticipate future advancements in the realm of physical attacks against infrared detectors.

\section{Related works}

\subsection{Physical attacks in the visible light field}

Physical attacks in the visible light field are broadly categorized into three main types: patch-based attacks, camera-based attacks, and light-based attacks.

\textbf{Patch-based attacks.} Patch-based attacks involve affixing printed adversarial patches onto the target object's surface to deceive advanced DNNs. In the context of face recognition, some studies \cite{ref18,ref19,ref20} have implemented attacks using carefully designed patches, employing covert strategies such as pasting cartoon patterns on faces to make them less suspicious to human observers. Adversarial patterns have also been placed on glasses or hats to reduce visibility to humans. Similarly, in pedestrian detection, works \cite{ref21,ref22,ref23,ref24} deploy adversarial patterns on clothing, evolving from conspicuous to more natural patterns over time. Given the non-rigid nature of clothing, robust optimization techniques like EOT \cite{ref26} and TPS \cite{ref27} are commonly employed to transition from digital to physical perturbations. In vehicle detection, prevalent physical attack methods \cite{ref11,ref12,ref25} utilize 3D rendering techniques to perturb various facets of the vehicle. To achieve robust adversarial effects, these methods often generate perturbations covering the entire vehicle, leveraging techniques like EOT and TV \cite{ref28} for the transition from the digital to the physical domain.

\textbf{Camera-based attacks.} Camera-based attacks revolve around introducing physical perturbations to the camera lens to deceive advanced DNNs. Li et al. \cite{ref29} pioneered AdvCS, initiating early exploration into camera-based attacks. This approach targets advanced DNNs by strategically affixing intricately designed patches onto camera lenses. Experimental results underscore the effectiveness of this method. Building on this foundation, Zolfi et al. \cite{ref30} enhanced AdvCS by introducing carefully designed translucent patches. This advancement involves creating patches with translucent properties to augment stealthiness. Simultaneously, their method refines the optimization function, facilitating the implementation of general adversarial attacks. Notably, this work strikes a balance between inconspicuousness and broader applicability.

\textbf{Light-based attacks.} Light-based attacks leverage light beams as physical perturbations projected onto the surface of a target object to deceive advanced DNNs. Duan et al. \cite{ref31} introduced AdvLB, employing random search optimization and laser beams as physical perturbations for attacks against advanced DNNs. Experimental results highlight the method's effectiveness in low-light environments. Hu et al. \cite{ref32} proposed AdvLS, utilizing laser points as physical perturbations and applying genetic algorithms \cite{ref33} for adversarial optimization, achieving both effectiveness and covertness in physical attacks. Zhong et al. \cite{ref34} introduced the shadow attack, utilizing particle swarm optimization \cite{ref38} and employing shadows as physical perturbations for natural and covert physical attacks. Experimental verification confirms its effectiveness in well-lit environments. Hu et al. \cite{ref35,ref36} investigated spotlight beams and color projection as physical perturbations for physical attacks against advanced DNNs. They employed stochastic optimization and particle swarm optimization for optimization, demonstrating the method's black-box adversarial impact in both digital and physical environments.

\subsection{Physical attacks in the infrared field}

Zhu et al. \cite{ref13} pioneered the exploration of infrared physical attacks with the bulb attack. This approach involves fitting a perturbation pattern using a Gaussian function and employing white-box optimization to generate adversarial examples. In the physical realm, lightbulb boards are utilized for implementing physical attacks against infrared pedestrian detectors. Experiments validate the method's effectiveness at various distances. Zhu et al. \cite{ref14} introduced the QR attack, focusing on designing a basic pattern with periodic expansion, ensuring the pattern maintains adversarial effects after random cropping and deformation. Deploying this basic pattern in the physical environment with aerogel as a thermal insulation material enables multi-view attacks. Wei et al. \cite{ref15} proposed AIP, leveraging aerogel as a thermal insulation material. Adversarial patches are generated through aggregation optimization, and in the physical environment, aerogel is cut to the shape of adversarial patches and deployed on pedestrians for physical attacks against infrared detectors. These three methods constitute white-box attacks, requiring knowledge of the target model's internal information, which is impractical in real-life scenarios. Only two existing infrared physical attacks are suitable for black-box scenarios. Wei et al. \cite{ref16} investigated HCB, utilizing cold and hot patches as physical perturbations. Modeled with a nine-perturbation grid, particle swarm optimization algorithm \cite{ref38} optimizes the position and shape of perturbations to generate effective and inconspicuous adversarial samples. Hu et al. \cite{ref17} introduced AdvIB, employing cold and hot stickers as perturbations and employing discrete disturbance modeling. The differential evolution algorithm \cite{ref47} optimizes the position and angle of perturbations for covert attacks against infrared pedestrian detectors. However, HCB and AdvIB struggle to strike a balance between stealthiness and robustness. Our proposed method utilizes cold patches, employs Bezier curves to model infrared perturbations, and uses PSO for optimization to generate the most adversarial Bezier curves, achieving a stealthy and robust physical attack against infrared pedestrian detectors.

\begin{figure*}
\centering
\includegraphics[width=0.7\linewidth]{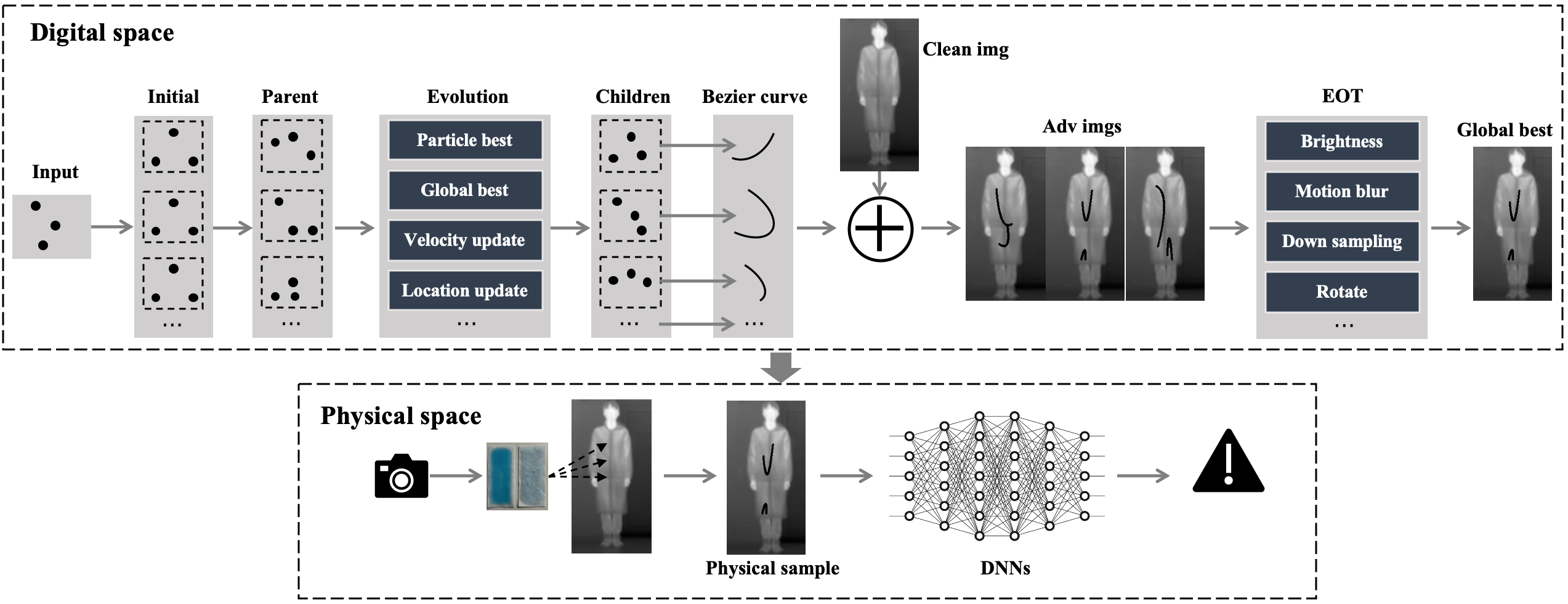} 
\caption{Graphical Summary. AdvIC generates adversarial simulation samples in a digital environment and deploys corresponding physical perturbations in the physical environment to create physical samples.}
\vspace{-0.3cm}
\label{figure2}
\end{figure*}

\section{Methodology}

\subsection{Problem definition}

\textbf{Object Detector:} In the context of a dataset $D$ containing infrared images of pedestrians, where $X$ and $Y$ represent the clean infrared images and ground truth labels for pedestrians, respectively, and $f$ denotes the object detector. Consequently, for each input image $X \in D$, the pretrained model of the object detector, denoted as $f: X \rightarrow Y$, predicts a label $y$ that aligns with the correct label $Y$. This label $y$ encompasses information about the bounding box position (${y}_{pos}$), object probability (${y}_{obj}$), and object class (${y}_{cls}$):

\begin{equation}
    \label{Formula 1}
    y:=[{y}_{pos},{y}_{obj},{y}_{cls}]=f(X)
\end{equation}

\textbf{Bezier Curve:} A Bezier curve serves as a mathematical representation employed for delineating smooth, curved paths within graphic design and computer graphics. Characterized by its definition through a set of control points, the positions and weights of these points dictate the curve's shape. Bezier curves of varying orders exhibit different quantities of control points; for instance, a first-order Bezier curve comprises two control points, a second-order Bezier curve includes three, and so forth. The mathematical formulation of Bezier curves empowers designers to effortlessly alter the curve's shape by manipulating the control points, offering a pliable and intuitive tool. Its mathematical formula is articulated as follows:

\begin{equation}
    \label{Formula 2}
    {B}_{n}(t)=\sum_{m=0}^{n} \binom{n}{m}{P}_{m}{(1-t)}^{n-i}{t}^{i}, \quad t \in [0,1]
\end{equation}
where ${B}_{n}(t)$ denotes the Bezier curve of order $n$.

\textbf{Particle Swarm Optimization (PSO):} PSO \cite{ref38} is a swarm intelligence-based optimization algorithm inspired by the collective behavior observed in groups of organisms such as bird flocks or fish schools. Introduced by computer scientists Russell Eberhart and James Kennedy in 1995, PSO offers key advantages:
(1) Global Search Capability. PSO excels in globally searching the problem space. It leverages collective cooperation and information sharing to discover global optimal solutions, avoiding local optima.
(2) Adaptability. PSO demonstrates strong adaptability to diverse problem topologies and characteristics, ensuring effective performance across different problem types.
(3) Versatility. PSO is applicable not only to continuous optimization problems but also to discrete optimization problems. This flexibility makes it a valuable tool for problem-solving in various domains.

\subsection{Generate adversarial sample}

Our approach is visually summarized in Figure \ref{figure2}, leveraging PSO for black-box optimization. Initially, discrete points are initialized and evolved, forming Bezier curves. Subsequently, the clean samples and Bezier curves undergo linear fusion to produce digital samples. During the digital-to-physical transition, robust optimization is executed through the EOT architecture \cite{ref26}. Finally, the perturbation is deployed in the physical world to execute a black-box attack against the infrared pedestrian detector.

\begin{figure}
\centering
\includegraphics[width=1\columnwidth]{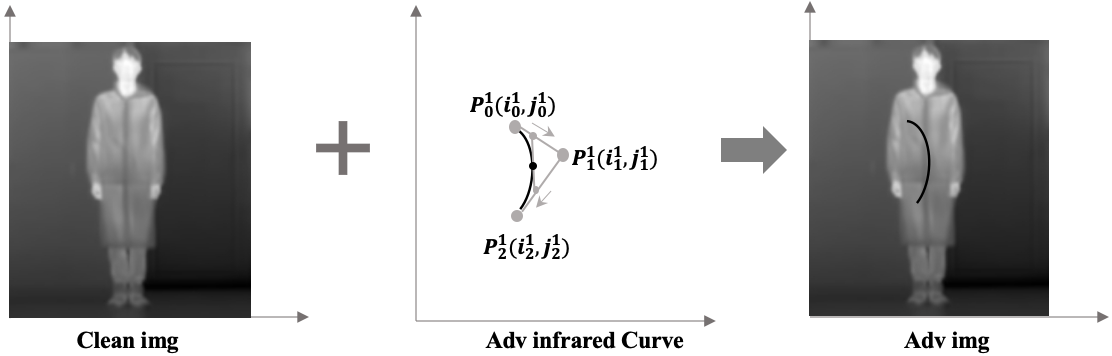} 
\caption{Schematic diagram of Bezier curve simulation modeling.}
\vspace{-0.3cm}
\label{figure3}
\end{figure}

In this study, we opt for the second-order Bezier curve for simulation modeling, illustrated in Figure \ref{figure3}. Each Bezier curve is defined by three vertices ${P}_{0}^{1}({i}_{0}^{1},{j}_{0}^{1})$, ${P}_{1}^{1}({i}_{1}^{1},{j}_{1}^{1})$, ${P}_{2}^{1}({i}_{2}^{1},{j}_{2}^{1})$, and is expressed as:

\begin{equation}
    \label{Formula 3}
    {B}_{2}^{1}(t)= {(1-t)}^{2}{P}_{0}^{1}+2t(1-t){P}_{1}^{1}+{t}^{2}{P}_{2}^{1}, \quad t \in [0,1]
\end{equation}
where ${B}_{2}^{1}(t)$ represents a second-order Bezier curve, and we denote multiple Bezier curves as $B(k)=\{{B}_{2}^{1}(t),{B}_{2}^{2}(t),...,{B}_{2}^{n}(t)\}$, where $k$ represents the number of second-order Bezier curves.

We use the cold patch as the physical perturbation in this study. Consequently, the modeled Bezier curve is set to the black state to simulate the perturbation pattern displayed by the cold patch under the infrared camera. We employ the linear fusion method $S$ to synthesize Bezier curves with clean samples, generating digital adversarial examples:

\begin{equation}
    \label{Formula 4}
    {X}_{adv}=S(X,B(k),\mathcal{M})
\end{equation}
where ${X}_{adv}$ represents the digital adversarial sample, and $\mathcal{M}$ represents the mask used to restrict the area of the Bezier curve, ensuring it does not extend beyond the pedestrain area in the image.

After creating a digital adversarial sample, we need to consider its transformation from the digital to the physical domain. To tackle this, we employ the Expectation Over Transformation (EOT) framework \cite{ref26}, known for effectively handling domain transfer challenges, especially in optimizing robustness for physical attacks. The framework utilizes a distribution of transformations $\mathcal{T}$, where each instance represents a combination of various random image transformations, such as viewpoint adjustments, brightness changes, downsampling, etc. Notably, $\mathcal{T}$ adeptly deals with slight errors in color and position of the simulated infrared perturbations. Thus, adversarial examples in the physical domain can be expressed as follows:

\begin{equation}
    \label{Formula 5}
    {X}_{adv} = {\mathbb{E}}_{t \sim \mathcal{T}}t({X}_{adv})
\end{equation}

\subsection{Infrared curves adversarial attack}

Our objective is to optimize the Bezier curve's shape using PSO to achieve the most adversarial configuration, which is then fused with the clean sample to generate the most adversarial digital sample. In this scenario, we address a more realistic black-box attack, where the attacker lacks internal model information and only has access to output results, including the detected target category and confidence. Consequently, we treat the target's confidence as an adversarial loss, and the formalized objective function in this work aims to minimize the infrared detector's confidence in the target object:

\begin{equation}
    \label{Formula 6}
    \mathop{\arg\min}_{\theta}{\mathbb{E}}_{t \sim \mathcal{T}}({y}_{obj} \leftarrow f(t({X}_{adv})))
\end{equation}

We utilize PSO for global optimization to obtain the most adversarial Bezier curves, which is then deployed in the physical world to deceive the infrared detector. The PSO optimization process involves the following steps:

\textbf{Initialization.} The PSO process begins by randomly generating a population of candidate solutions, comprising the population ($POP$) and their associated velocity vector ($V$):

\begin{equation}
    \label{Formula 7}
    POP=[{B}_{1}(k),{B}_{2}(k),...,{B}_{\alpha}(k)]
\end{equation}

\begin{equation}
    \label{Formula 8}
    V=[{v}_{1},{v}_{2},...,{v}_{\alpha}]
\end{equation}
where $\alpha$ represents the population size, ${B}_{a}(k)$ denotes each candidate solution in the population POP, and $a$ ranges from 1 to $\alpha$. ${v}_{a}$ represents the direction of movement for particle ${B}_{a}(k)$.

\textbf{Generate adversarial examples.} After random initialization, adversarial samples are generated for each individual ${B}_{a}^{b}(k)$ in the population POP using equation \ref{Formula 4}:

\begin{equation}
    \label{Formula 9}
    {X}_{a}^{b}=S(X,{B}_{a}^{b}(k),\mathcal{M})
\end{equation}
where $b$ represents the current iteration number of the population. ${X}_{a}^{b}$ denotes the adversarial sample generated by the $a$-th individual in the population of generation $b$.

\textbf{Obtain the individual and global optimum.} To guide the evolution of the population towards the optimal solution, we need to determine the individual optimal solution for each particle and the global optimal solution for the entire population:

\begin{equation}
    \label{Formula10}
    {B}_{a,best}^{b}(k)=\mathop{\arg\min}_{{B}_{a,best}^{u}(k)}{y}_{obj} \leftarrow f({X}_{a}^{u})  \quad u \in [1, b]
\end{equation}

\begin{equation}
    \label{Formula11}
    {B}_{best}^{b}(k)=\mathop{\arg\min}_{{B}_{a,best}^{u}(k)}{y}_{obj} \leftarrow f({X}_{a}^{u}) \quad a \in [1, \alpha] , u \in [1, b]
\end{equation}
where ${B}_{a,best}^{b}(k)$ and ${B}_{best}^{b}(k)$ represent the optimal solution of the $a$-th individual in the $b$-generation population $POP$ and the global optimal solution of the population, respectively.

\textbf{Update the velocity and position information of the individual.} After obtaining the individual optimal solution and the global optimal solution, the information is used to guide the population towards the direction that is easier to obtain the optimal solution. The following formula is used to update the individual velocity direction and position of the population:

\begin{equation}
    \label{Formula12}
    {v}_{a}^{b+1}=\omega{v}_{a}^{b}+{c}_{1}{r}_{1}({B}_{a,best}^{b}(k)-{B}_{a}^{b}(k)) + {c}_{2}{r}_{2}({B}_{best}^{b}(k)-{B}_{a}^{b}(k))
\end{equation}

\begin{equation}
    \label{Formula13}
    {B}_{a}^{b+1}(k)={B}_{a}^{b}(k)+{v}_{a}^{b+1}
\end{equation}
where $\omega$, ${c}_{1}$, ${r}_{1}$, ${c}_{2}$, ${r}_{2}$, are the hyperparameters of PSO, and $\omega$ denotes the inertia factor. ${c}_{1}$ and ${c}_{2}$ denote the learning factors of the particles. ${r}_{1}$ and ${r}_{2}$ are random numbers generated from a uniform distribution in the range [0,1].

Algorithm \ref{algorithm1} provides the pseudo-code for the proposed TOUAP. It takes the clean sample $X$, object detector $f$, population size $\alpha$, iteration number $I$, and PSO hyperparameters $\omega$, ${c}_{1}$, ${r}_{1}$, ${c}_{2}$, ${r}_{2}$ as input. The algorithm's detailed optimization process is delineated in Algorithm 1, and the output is the optimal solution of the population after the $I$-th iteration. This optimal solution represents the most adversarial Bezier curve, utilized in subsequent physical attack experiments.

\begin{algorithm}
	\renewcommand{\algorithmicrequire}{\textbf{Input:}}
	\renewcommand{\algorithmicensure}{\textbf{Output:}}
	\caption{Pseudocode of AdvCL}
	\label{algorithm1}
	\begin{algorithmic}[1]
	
		\REQUIRE Clean sample $X$, Detector $f$, Population size $\alpha$, Iterations $I$, Hyperparameters of PSO: $\omega$,${c}_{1}$,${r}_{1}$,${c}_{2}$, ${r}_{2}$;
		\ENSURE Physical parameters ${B}^{\star}$;

		\STATE \textbf{Initialization} Randomly set $POP$, $V$;
        \FOR{$b$ $\leftarrow$ 0 to $I$}
            \FOR{each ${B}_{a}^{b}(k)$ in ${POP}^{b}$}
                \STATE ${X}_{a}^{b}=S(X,{B}_{a}^{b}(k),\mathcal{M})$;
                \STATE ${B}_{a,best}^{b}(k)=\mathop{\arg\min}_{{B}_{a,best}^{u}(k)}{y}_{obj} \leftarrow f({X}_{a}^{u})  \quad u \in [1, b]$;
                \STATE ${B}_{best}^{b}(k)=\mathop{\arg\min}_{{B}_{a,best}^{u}(k)}{y}_{obj} \leftarrow f({X}_{a}^{u}) \quad a \in [1, \alpha] , u \in [1, b]$;
                \STATE ${B}^{\star}={B}_{best}^{b}(k)$;      
            \ENDFOR
            
            \STATE ${v}_{a}^{b+1}=\omega{v}_{a}^{b}+{c}_{1}{r}_{1}({B}_{a,best}^{b}(k)-{B}_{a}^{b}(k)) + {c}_{2}{r}_{2}({B}_{best}^{b}(k)-{B}_{a}^{b}(k))$;
            \STATE ${B}_{a}^{b+1}(k)={B}_{a}^{b}(k)+{v}_{a}^{b+1}$;
        \ENDFOR
        
        \STATE \textbf{Output:} ${B}^{\star}$;

	\end{algorithmic}  
\end{algorithm}

\section{Experiments}
\label{sec4}

\subsection{Experimental setting}

\textbf{Dataset:} In line with AdvIB \cite{ref17}, we employ the FLIR dataset \cite{ref39} for both training the infrared pedestrian detector model and conducting digital attack tests. The FLIR dataset comprises 10,228 infrared images captured by the FLIR Tau2 thermal imager. To prepare the dataset for model training, we filter pedestrians with a height greater than 120 pixels, resulting in a subset of 1,011 images. Model training is then executed on this dataset, with subsequent digital attacks performed on its test set.

\textbf{Object Detector:} Similar to AdvIB, this study employs YOLOv3 \cite{ref4}, DETR \cite{ref41}, Mask R-CNN \cite{ref42}, Faster R-CNN \cite{ref43}, Libra R-CNN \cite{ref44}, and RetinaNet \cite{ref45} as object detectors, all trained on the filtered dataset. The resulting average accuracies on the test set are 90.7\%, 91.2\%, 89.5\%, 90.8\%, 88.0\%, and 93.0\%, respectively.

\begin{figure}
\centering
\includegraphics[width=1\columnwidth]{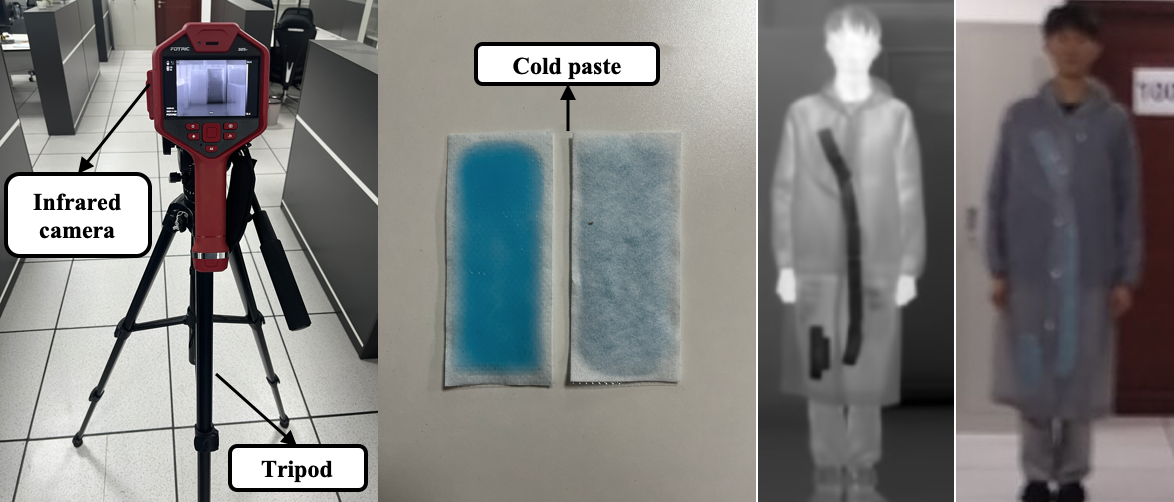} 
\caption{Experimental devices. The experimental devices for AdvIC include an infrared camera, a tripod, and cold pastes.}
\vspace{-0.3cm}
\label{figure4}
\end{figure}

\textbf{Experimental devices:} In the physical attack experiments, our devices, depicted in Figure \ref{figure4}, comprise a tripod, an infrared camera, and cold patches. The infrared camera used is the FOTRIC 325+, featuring parameters such as FPA of 352×264, NETD less than 40mk, and SR of 704×528. AdvIC is tested across various infrared cameras, demonstrating consistent effectiveness. The cold patches used can maintain a temperature of ${24}^{\circ}C$ for ten hours. Figure \ref{figure4} illustrates the perturbation effects of the cold patch captured under both the infrared and visible cameras, highlighting the significant impact visible under the infrared camera.

\textbf{Evaluation Metrics.} The primary objective of our work is to achieve a disappearance attack, rendering the object detector unable to detect pedestrians after introducing the perturbation. Hence, we employ the Attack Success Rate (ASR) as the key metric to assess the adversarial impact of AdvIC. A higher ASR indicates a more effective adversarial impact, while a lower ASR suggests a less successful attack. The ASR is calculated using the following formula:

\begin{equation}
\label{eq:Positional Encoding}
\begin{split}
    &{\rm ASR}(X) = 1-\frac{1}{G}\sum_{g=1}^{G}F({y}_{obj}^{g})\\
    &F({y}_{obj}^{g})=
        \begin{cases}
        0 & {y}_{obj}^{g} < 0.5 \\
        1 & otherwise
        \end{cases}
\end{split}
\end{equation}
where $G$ represents the number of true positive labels in the dataset that the detector can detect without any attacks. For all our attack experiments, we maintain a threshold of 0.5. In other words, the disappearance attack is considered successful when the confidence of the detected target falls below this threshold.

\textbf{Competing methods.} The proposed TOUAP falls under black-box attacks, and for a fair comparison, we select black-box attacks HCB \cite{ref16} and AdvIB \cite{ref17} as baselines for experimental comparison.

\textbf{Other details.} We set the hyperparameters of PSO as follows: $\omega=0.9$, ${c}_{1}=1.6$, ${r}_{1}=0.5$, ${c}_{2}=1.4$, ${r}_{2}=0.5$. All attack experiments are conducted on an NVIDIA GeForce RTX 4090 GPU.

\subsection{Evaluation of effectiveness}

\begin{figure}
\centering
\includegraphics[width=1\columnwidth]{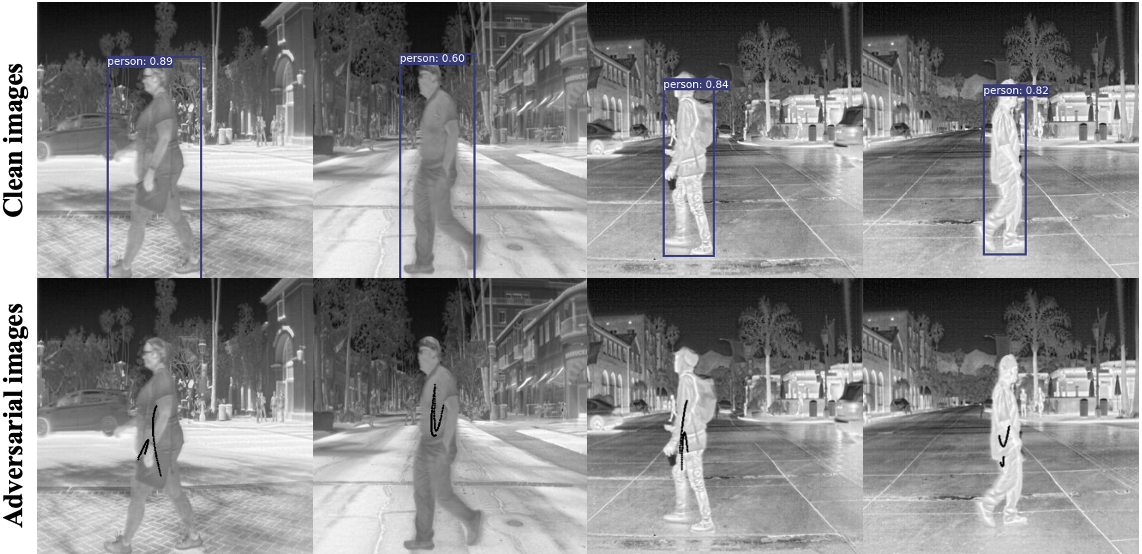} 
\caption{Digital Samples. The first row displays the detection results of clean samples, while the second row shows the detection results of adversarial samples with two Bezier curves.}
\vspace{-0.3cm}
\label{figure5}
\end{figure}

\textbf{Digital attacks.} We conduct a series of ablation experiments to assess the effectiveness of the proposed AdvIC in the digital domain, while selecting reasonable physical attack configurations based on these experimental results. Given that physical perturbations can only represent black or white patterns in the thermal imager, we choose white and black Bezier curves as perturbations to perform the digital attack against Yolo v3. We set the number of Bezier curves ($k$) from 1 to 7, and the results of our digital attack experiments are shown in Table \ref{Table1}. We can draw the following conclusions: \textbf{1)} our method achieves excellent adversarial effects in the digital domain, achieving a success rate of 88.51\% even using only a black Bezier curve; \textbf{2)} The success rate of AdvIC increases with the number of Bezier curves, which is consistent with our expectation; \textbf{3)} The adversarial effect of a black Bezier curve is better than that of white, so we choose black as the perturbation in the physical experiment; \textbf{4)} When the number of black Bezier curves reaches 2, the growth rate of AdvIC's attack success rate slows down, and the attack success rate reaches 94.83\%, so we select 2 Bezier curves for the physical experiment. In subsequent experiments, we use the attack configuration (two black Bezier curves) for digital/physical attack experiments. Note that we did not perform the cross-color digital attack experiment, this is because the white Bezier curve is weak in the digital environment, and when the hot patch is pasted on the pedestrian's clothing, its width is too large (the hot patch cannot be cut), resulting in a large physical perturbation. Figure \ref{figure5} shows the adversarial samples generated by our method in the digital environment, and it can be seen that the perturbations generated by our method exhibit better stealthiness.

\begin{table} 
	\centering
 
    \setlength{\belowcaptionskip}{5pt}
    \caption{Results of ablation experiments for color and $k$.}
    \label{Table1}
    \resizebox{\columnwidth}{!}{
	\begin{tabular}{ccccccccc}

    \hline
    ~&~&$k=$1&$k=$2&$k=$3&$k=$4&$k=$5&$k=$6&$k=$7\\
    \hline
    \multirow{2}*{Black}&ASR&88.5&94.8&96.6&96.6&96.6&96.6&96.0\\
    \cmidrule(r){2-9}
    ~&Query&114.8&67.8&45.2&37.7&38.5&36.8&36.0\\
    \hline
    \multirow{2}*{White}&ASR&63.2&78.2&79.9&82.8&83.3&82.2&83.9\\
    \cmidrule(r){2-9}
    ~&Query&237.4&182.6&165.2&153.3&145.8&142.0&143.0\\
    \hline

\end{tabular}
}
\end{table}

\begin{figure*}
\centering
\includegraphics[width=1\linewidth]{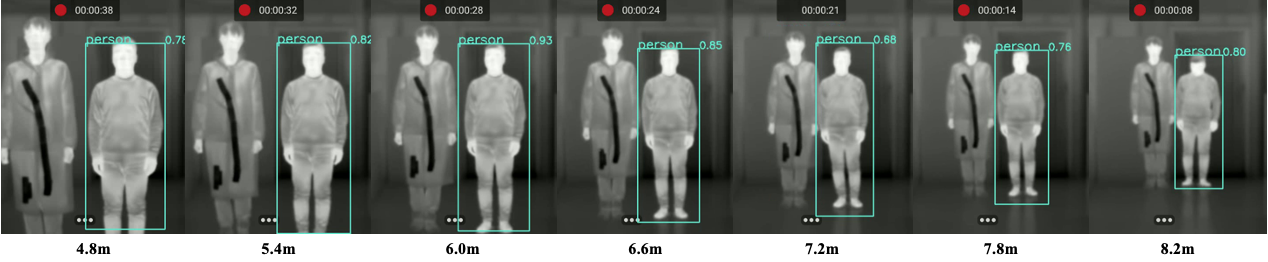} 
\caption{Physical samples.}
\vspace{-0.3cm}
\label{figure6}
\end{figure*}

\textbf{Physical Attacks.} To comprehensively assess the adversarial efficacy of AdvIC, we conducted physical experiments targeting Yolo v3 at different distances, ranging from 4.8m to 8.4m, with a 0.6m interval. Employing a predefined attack configuration, the results are detailed in Table \ref{Table2}. AdvIC consistently achieves effective physical attacks across all distances, reaching a 100\% attack success rate at a distance of 7.8 meters. Figure \ref{figure6} visually represents the generated physical samples, highlighting the subtle nature of the perturbations. Notably, only two Bezier curves suffice to successfully deceive the infrared pedestrian detector across varying distances.

\begin{table} 
	\centering
 
    \setlength{\belowcaptionskip}{5pt}
    \caption{Experimental results of AdvIC physical attacks against Yolo v3 at different distances.}
    \label{Table2}
    \resizebox{\columnwidth}{!}{
	\begin{tabular}{ccccccccc}

    \hline
    Distance&4.8m&5.4m&6.0m&6.6m&7.2m&7.8m&8.4m&avg\\
    \hline
    ASR&58.4&28.3&61.4&56.6&82.7&100&79.4&67.2\\
    \hline

\end{tabular}
}
\end{table}

In summary, the results presented in Table \ref{Table1} confirm the effectiveness of AdvIC in the digital domain, aiding in the selection of a suitable attack configuration. Additionally, the outcomes in Table \ref{Table2} demonstrate the effectiveness of AdvIC in the physical domain.

\subsection{Evaluation of stealthiness}

\begin{figure}
\centering
\includegraphics[width=1\columnwidth]{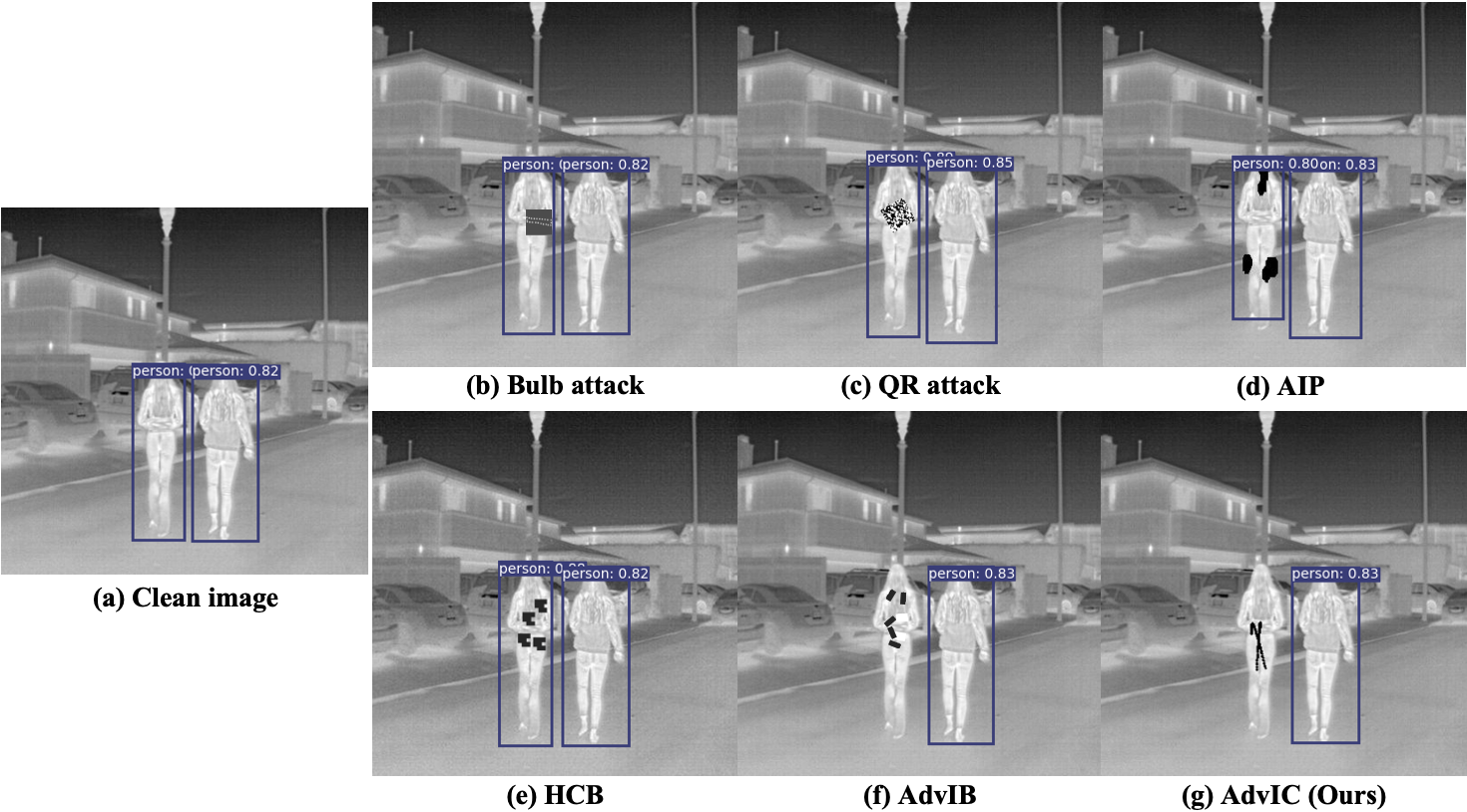} 
\caption{Comparison and presentation of digital adversarial samples.}
\vspace{-0.3cm}
\label{figure7}
\end{figure}

In the digital environment, as depicted in Figure \ref{figure7}, our method generates digital samples that exhibit a more natural and inconspicuous visual effect compared to the baseline method. This, combined with the presentation of digital samples in Figure \ref{Formula 5}, highlights the enhanced stealthiness of AdvIC in the digital domain when compared to the baseline method. In the physical domain, as illustrated in Figure \ref{Formula 4}, our approach involves pasting the cold patch inside the clothing, making its presence challenging to detect without the aid of an infrared camera. Consequently, our method demonstrates greater covert characteristics compared to the bulb attack \cite{ref13}, QR attack \cite{ref14}, and AIP \cite{ref15}, where perturbations are deployed externally. Figure \ref{figure1} (c), (d), and (e) further reveal that the physical samples generated by AdvIC exhibit a more natural curve texture under the infrared camera, appearing smoother and more concealed than HCB \cite{ref16} and AdvIB \cite{ref17}. Through these observations, we establish the stealthiness of AdvIC in both digital and physical domains.

\subsection{Evaluation of robustness}

We subjected AdvIC to various advanced object detectors, namely DETR \cite{ref41}, Mask Rcnn \cite{ref42}, Faster Rcnn \cite{ref43}, Libra Rcnn \cite{ref44}, and RetinaNet \cite{ref45}[45], in order to assess its robustness. The summarized experimental results are presented in Table \ref{Table3}, leading to the following conclusions: \textbf{1)} AdvIC attains an average attack success rate of 76.8\% against advanced detectors, affirming the effectiveness of our method in targeting diverse advanced detectors; \textbf{2)} The average query times for AdvIC with various advanced detectors amount to 159.0, highlighting the efficiency of our proposed method; \textbf{3)} DETR exhibits a more resilient detection capability compared to other detectors when faced with AdvIC attacks. The attack success rate against DETR is only 33.3\%, potentially indicating the superior robustness of Transformer-based detectors.

\begin{table} 
	\centering
 
    \setlength{\belowcaptionskip}{5pt}
    \caption{Experimental results of deploying AdvIC to attack various advanced detectors.}
    \label{Table3}
    \resizebox{\columnwidth}{!}{
	\begin{tabular}{cccccccc}

    \hline
    $f$&Yolo v3&DETR&Mask&Faster&Libra&Retina&avg\\
    \hline
    ASR&94.8&33.3&79.4&92.7&93.0&67.8&76.8\\
    \hline
    Query&67.8&366.8&162.0&86.2&66.9&204.7&159.0\\
    \hline

\end{tabular}
}
\end{table}

\begin{table} 
	\centering
 
    \setlength{\belowcaptionskip}{5pt}
    \caption{Comparison of experimental results between the proposed AdvIC and baseline methods.}
    \label{Table4}
	\begin{tabular}{cccc}

    \hline
    Method&HCB&AdvIB&AdvIC\\
    \hline
    ASR&37.3&56.7&\textbf{76.8}\\
    \hline
    Query&694.0&472.8&\textbf{159.0}\\
    \hline

\end{tabular}
\end{table}

We present the experimental outcomes of AdvIC and the baseline method against different advanced detectors in Table \ref{Table4}. It is evident that our method achieves a higher average attack success rate and lower average query times than the baseline methods, substantiating that AdvIC is more effective in generating adversarial effects. 

The results in Table \ref{Table3} demonstrate that AdvIC can effectively attack various advanced detectors, showcasing its generalization ability. The findings in Table \ref{Table4} further emphasize the heightened adversarial efficacy of AdvIC compared to the baseline methods. Consequently, the combined results from Tables \ref{Table3} and \ref{Table4} provide strong evidence for the robustness of AdvIC.

\section{Discussion}

\subsection{Transferability of AdvIC}

To assess the transfer attack capability of AdvIC in the digital domain, we utilize samples generated by the digital attack, which successfully attacked Yolo v3, as a dataset for conducting transfer attacks against DETR, Mask Rcnn, Faster Rcnn, Libra Rcnn, and RetinaNet. The transfer ASR achieved are 3.85\%, 22.89\%, 29.07\%, 29.70\%, and 26.45\%, respectively.

In the physical domain, we employ the physical samples that successfully attacked Yolo v3 at various distances as a dataset to conduct transfer attacks against DETR, Mask Rcnn, Faster Rcnn, Libra Rcnn, and RetinaNet. The experimental results are summarized in Table \ref{Table5}, leading to the following conclusions: \textbf{1)} Generally, physical samples generated by AdvIC exhibit effective transfer attack effects against different advanced detectors. \textbf{2)} For the transfer attack from Yolo v3 to DETR, the adversarial effect becomes more pronounced with larger distances, achieving a 100\% success rate in physical transfer attacks at 7.8 meters. Conversely, the physical transfer against DETR exhibits a weaker effect at smaller distances. \textbf{3)} The transfer attack from Yolo v3 to Mask Rcnn achieves no less than a 50\% physical migration success rate at each distance, indicating effective transfer attacks from Yolo v3 to Mask Rcnn at various distances. \textbf{4)} For the transfer attack from Yolo v3 to Faster Rcnn, a significant adversarial transfer effect is observed at smaller distances, with the effect weakening as the distance increases. \textbf{5)} In the transfer attack from Yolo v3 to Libra Rcnn, the adversarial effect is weak at a medium distance, achieving only a 9.3\% transfer success rate at 6.6 meters. However, the adversarial effect becomes more pronounced at smaller or larger distances. \textbf{6)} In the Yolo v3 to RetinaNet transfer attack, a significant adversarial transfer effect is achieved at all distances, reaching a 100\% physical transfer success rate at distances of 6.6 meters and 8.4 meters.

In summary, AdvIC demonstrates relatively weak digital transfer attack effects, while in the physical domain, it achieves significant transfer effects in most cases. This observation suggests that the dataset used for digital attacks contains diverse infrared pedestrians with complex background environments, making robust transfer attacks challenging. However, in the physical transfer attack experiment, the infrared pedestrians in the samples are relatively uniform, and the background is simple, facilitating stable physical transfer attacks.

\begin{table} 
	\centering
 
    \setlength{\belowcaptionskip}{5pt}
    \caption{Experimental results of physical transfer attacks.}
    \label{Table5}
	\begin{tabular}{cccccccc}

    \hline
    $f$&4.8m&5.4m&6.0m&6.6m&7.2m&7.8m&8.4m\\
    \hline
    DETR&14.3&2.7&54.6&86.1&89.4&100.0&75.5\\
    \hline
    Mask&84.4&64.9&56.5&53.5&91.1&83.5&70.9\\
    \hline
    Faster&85.7&94.6&98.2&67.4&78.1&35.1&51.0\\
    \hline
    Libra&71.4&59.5&68.5&9.3&15.5&19.6&70.9\\
    \hline
    Retina&55.8&78.4&95.4&100.0&98.4&79.4&100.0\\
    \hline

\end{tabular}
\end{table}

\subsection{Ablation study of shape}

\begin{figure}
\centering
\includegraphics[width=1\columnwidth]{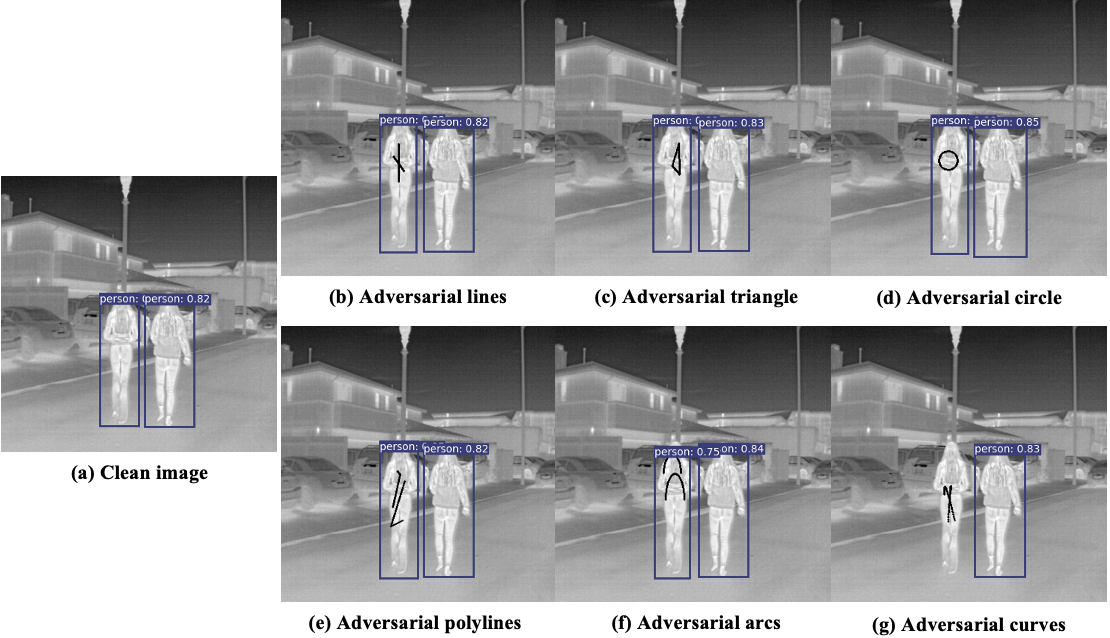} 
\caption{Demonstration of digital samples generated by different-shape infrared perturbations.}
\vspace{-0.3cm}
\label{figure8}
\end{figure}

In this study, the Bezier curve is chosen to simulate the adversarial infrared curve, and to assess the superiority of the Bezier curve, a series of comparative experiments are conducted. Illustrated in Figure \ref{figure8} are perturbation shapes consisting of two straight lines (Figure \ref{figure8} (b)), a triangle (Figure \ref{figure8} (c)), a circle (Figure \ref{figure8} (d)), two polylines (Figure \ref{figure8} (e)), and two 180-degree circular arcs (Figure \ref{figure8} (f)). The digital ablation experimental results on the shapes attacking Yolo v3 are summarized in Table \ref{Table6}. The findings reveal that the use of various shapes of infrared perturbations can achieve significant adversarial effects against Yolo v3. Notably, the use of the Bezier curve demonstrates the most pronounced adversarial effect, achieving an attack success rate of 94.8\%. Additionally, the adversarial effect of two polylines is lighter and weaker than that of two straight lines, which deviates from the expected outcome.

\begin{table} 
	\centering
 
    \setlength{\belowcaptionskip}{5pt}
    \caption{Ablation study of shape.}
    \label{Table6}
	\begin{tabular}{ccccccc}

    \hline
    Shape&Lines&Triangle&Circle&Polylines&Arcs&Ours\\
    \hline
    ASR&76.1&83.3&97.2&74.4&87.8&\textbf{94.8}\\
    \hline
    Query&151.8&114.6&21.2&159.5&83.0&\textbf{67.8}\\
    \hline

\end{tabular}
\end{table}

\subsection{Deploy AdvIC to attack infrared vehicle detectors}

In addition to verifying the adversarial effect of AdvIC against an infrared pedestrian detector, we extend its application to infrared vehicle detectors. These detectors are trained on the FLIR dataset, including Yolo v3 \cite{ref4}, DETR \cite{ref41}, Mask Rcnn \cite{ref42}, Faster Rcnn \cite{ref43}, Libra Rcnn \cite{ref44}, and RetinaNet \cite{ref45}, achieving average accuracies of 92.1\%, 94.6\%, 94.2\%, 94.4\%, 95.6\%, and 95.5\%, respectively. When deploying AdvIC to attack these meticulously trained infrared vehicle detectors, the experimental results are summarized in Table \ref{Table7}. It is evident that AdvIC can effectively induce adversarial effects on each infrared vehicle detector, achieving an average success rate of 42.5\%, with an average query count of 305.80. The most significant adversarial effect is observed when attacking Faster Rcnn, achieving an 80.5\% success rate.

\begin{table} 
	\centering
 
    \setlength{\belowcaptionskip}{5pt}
    \caption{Deploy AdvIC to attack infrared vehicle detectors.}
    \label{Table7}
    \resizebox{\columnwidth}{!}{
	\begin{tabular}{cccccccc}

    \hline
    $f$&Yolo v3&DETR&Mask&Faster&Libra&Retina&avg\\
    \hline
    ASR&36.6&35.4&40.2&80.5&38.5&24.1&42.5\\
    \hline
    Query&340.4&345.4&316.4&136.9&336.1&359.6&305.8\\
    \hline

\end{tabular}
}
\end{table}

\subsection{Defense of AdvIC}

In addition to presenting the attack performance of AdvIC, we conduct adversarial defense experiments against it. Following our investigation, we opt for adversarial training \cite{ref14} and DW \cite{ref46} as defense mechanisms to assess their efficacy. In adversarial training, we introduce randomly generated Bezier curves to clean samples to create adversarial samples. Subsequently, we augment the training dataset by incorporating both clean and adversarial samples in a 1:5 ratio to conduct adversarial training. This process results in an adversarially trained Yolo v3 detector with an average accuracy of 90.9\%. As for DW, it primarily includes two approaches. The first is blind image inpainting, where the defender lacks information about the perturbation's location, necessitating detection before performing image inpainting. The second is non-blind image inpainting. In this scenario, the defender has access to the position information of the perturbation in the adversarial sample and can perform targeted image inpainting. In this work, we adopt non-blind image inpainting. Specifically, upon successfully generating an adversarial example, we modified the pixel values in the perturbed region to (178, 178, 178), which is the closest value to the pixel values in the pedestrian region of the clean sample. The final experimental results of adversarial training and DW defense are presented in Table \ref{Table8}. It is evident that both adversarial training and DW achieve effective adversarial defense against AdvIC, reducing its attack success rate and increasing the query cost. However, neither adversarial training nor DW could achieve comprehensive defense against AdvIC.

\begin{table} 
	\centering
 
    \setlength{\belowcaptionskip}{5pt}
    \caption{Adversarial defense strategies against AdvIC.}
    \label{Table8}
	\begin{tabular}{cccc}

    \hline
    ~&No defense&Adversarial training&DW\\
    \hline
    ASR&94.8&39.6&48.3\\
    \hline
    Query&67.8&391.6&306.8\\
    \hline

\end{tabular}
\end{table}

\section{Conclusion}

In this study, we propose a novel black-box physical attack against infrared pedestrian detectors called Adversarial Infrared Curves (AdvIC). AdvIC utilizes Bezier curves to model infrared curves and employs PSO to optimize these curves for generating the most adversarial infrared patterns. The evaluation criteria for this work include effectiveness, stealthiness, and robustness.

Effectiveness is demonstrated by achieving attack success rates of 94.8\% and 67.2\% in digital and physical environments, respectively, highlighting the efficacy of AdvIC. Stealthiness is confirmed through comparisons between digital and physical samples generated by AdvIC and a baseline method, revealing AdvIC's superior imperceptibility. Regarding robustness, AdvIC is deployed against various advanced detectors, achieving an average ASR of 76.8\%, surpassing the baseline methods and validating its robustness.

This method introduces a new approach to infrared physical attacks by leveraging smooth curves to enhance the stealthiness of such attacks. In future work, we aim to explore physical attacks in the context of infrared-visible cross-modal and infrared-visible-radar cross-modal scenarios. Additionally, developing more effective defense mechanisms against infrared physical attacks remains a focus of our ongoing research.








\bibliographystyle{IEEEtran}
\bibliography{IEEEfull}


\end{document}